\documentstyle[psfig,12pt]{article}
\def\be{\begin{equation}}       
\def\ee{\end{equation}}
\def\bear{\be\begin{array}}      
\def\eear{\end{array}\ee}
\def\bea{\begin{eqnarray}}
\def\eea{\end{eqnarray}}

\def\etal{{\it et al.}}

\def\bold#1{\setbox0=\hbox{$#1$}%
     \kern-.025em\copy0\kern-\wd0
     \kern.05em\copy0\kern-\wd0
     \kern-.025em\raise.0433em\box0 }

\catcode`\@=11  

\@addtoreset{equation}{section}

\begin{document}

\baselineskip 18pt

\newcommand{\sheptitle}
{\Large Radiatively Corrected Chargino Pair Production at 
LEP2${}^{\dag}$}

\newcommand{\shepauthor}
{Marco A. D\'\i az${}^1$, Steve F. King${}^{2*}$, and Douglas A. Ross${}^3$}

\newcommand{\shepaddress}
{${}^1$ Departamento de F\'\i sica Te\'orica, IFIC--CSIC, Universidad de 
Valencia\\ 
Burjassot, Valencia 46100, Spain \\
${}^2$ CERN, Theory Division, CH--1211 Geneva, Switzerland\\
${}^3$ Department of Physics and Astronomy, University of Southampton\\
Southampton, SO17 1BJ, U.K.}

\newcommand{\shepabstract}
{
One--loop radiative corrections to the production cross section of a pair
of light charginos in $e^+e^-$ colliders are calculated within the MSSM. 
Top and bottom quarks and squarks are considered in the loops, and they 
are renormalized using the $\overline{MS}$ scheme. If the center of mass
energy is equal to 192 GeV, positive corrections typically of $10\%$ to 
$15\%$ are found when the squark mass parameters are equal to 1 TeV.
}

\begin{titlepage}
\begin{flushright}
CERN-TH/98-26\\
SHEP--98/02\\
FTUV/98--7\\
IFIC/98--7\\
hep-ph/9801373\\
\end{flushright}
\vspace{.1in}
\begin{center}
{\large{\bf \sheptitle}}
\bigskip \\ \bigskip\shepauthor\bigskip \\ 
\mbox{} \\ {\it \shepaddress} \\ \vspace{.5in}
{\bf Abstract} \bigskip \end{center} \setcounter{page}{0}
\shepabstract\\
\begin{flushleft}
CERN-TH/98-26\\
January 1998
\end{flushleft}

\noindent ${}^{\dag}$Talk given by M.A.D. at the International Workshop
``Beyond the Standard Model: From Theory to Experiment'', 
13--17 October 1997, Valencia, Spain.\\
\noindent${}^*$ On leave of absence from $^3$.

\end{titlepage}

\setcounter{page}{1}

In the Minimal Supersymmetric Standard Model (MSSM), the supersymmetric
partners of the charged Higgs and the $W$ gauge bosons mix to
form a set of two charged fermions called charginos.
Experimental searches for charginos at LEP2 have been negative so far, 
and lower bounds on the lightest chargino mass have been set. The bound 
depends mainly on the sneutrino mass and the mass difference between the 
chargino and the LSP $\Delta m=m_{\chi_1}-m_{\chi^0_1}$.
ALEPH has found that $m_{\chi_1}>85$ GeV for $m_{\tilde\nu_e}>200$ GeV 
\cite{ALEPH}. DELPHI's bound corresponds
to $m_{\chi_1}>84.3$ GeV for $m_{\tilde\nu_e}>300$ GeV and 
$\Delta m>10$ GeV \cite{DELPHI}. A lower bound of $m_{\chi_1}>85.5$ GeV
was found by L3 for $m_{\tilde\nu_e}>300$ GeV \cite{L3}.
Finally, OPAL has found that if $\Delta m>10$ GeV then 
$m_{\chi_1}>84.5$ GeV if $m_0>1$ TeV and $m_{\chi_1}>65.7$ GeV for 
the smallest $m_0$ compatible with current limits on sneutrino
and slepton masses \cite{OPAL}.

\begin{figure}
\centerline{\protect\hbox{\psfig{file=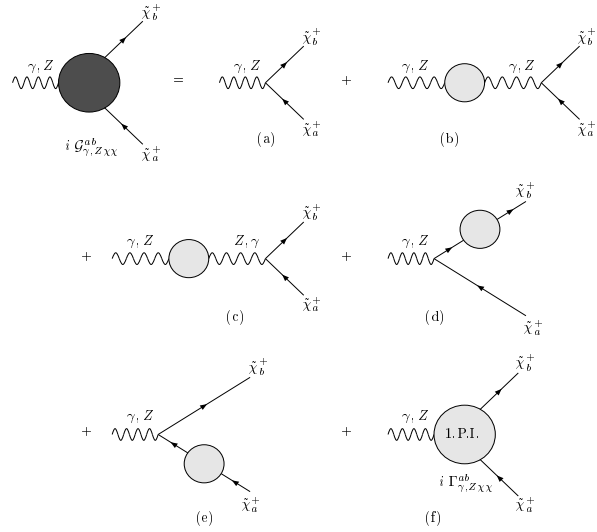,height=7cm,width=0.5\textwidth}}}
\caption{One--loop renormalized $\gamma\chi^+_b\chi^-_a$ and 
$Z\chi^+_b\chi^-_a$ vertex functions.} 
\label{GZchacha1lvertex} 
\end{figure} 

In the Born approximation, chargino masses and mixing angles in the MSSM
depend only on the $SU(2)$ gaugino mass $M$, the ratio between Higgs 
vacuum expectation values $\tan\beta$, and the supersymmetric Higgs 
mass parameter $\mu$. Much can be learned about these parameters from
an accurate measurement of the total chargino production cross section
and masses in $e^+e^-$ colliders, and also about the rest of the 
supersymmetric particles in the case of MSSM--SUGRA \cite{ChaParam}. 
Nevertheless, for this to work, accurate experimental measurements 
must be accompanied by precise theoretical calculations. In this talk
we report on a recent calculation of the one--loop corrections to the 
total production cross section of a pair of charginos in $e^+e^-$
colliders \cite{chargRC}.

Charginos are produced in the s--channel with intermediate $Z$ bosons and 
photons, and in the t--channel with intermediate electron sneutrino. In our
approximation only $Z\tilde\chi^+\tilde\chi^-$, 
$\gamma\tilde\chi^+\tilde\chi^-$, $e^+\tilde\nu_e\tilde\chi^-$, and 
$e^-\tilde\nu_e\tilde\chi^+$ vertices are renormalized. We denote
these one--loop renormalized total vertex functions
$i{\cal G}_{Z\chi\chi}^{ab}$, $i{\cal G}_{\gamma\chi\chi}^{ab}$, 
$i{\cal G}_{\tilde{\nu_e}e\chi}^{+b}$, and
$i{\cal G}_{\tilde{\nu_e}e\chi}^{-a}$ respectively. 
The first two total vertex functions are given in 
Fig.~\ref{GZchacha1lvertex}, where we have the following contributions: 
(a) tree level, (b) gauge boson self energies, (c) $Z-\gamma$ mixing, 
(d)--(e) chargino self energy and chargino mixing, and (f) the 1PI
triangular diagrams. The two total vertex functions involving sneutrinos 
are simpler because they receive contributions only from the tree level 
vertex and chargino self energy and mixing, and we do not display them.

\begin{figure}
\centerline{\protect\hbox{\psfig{file=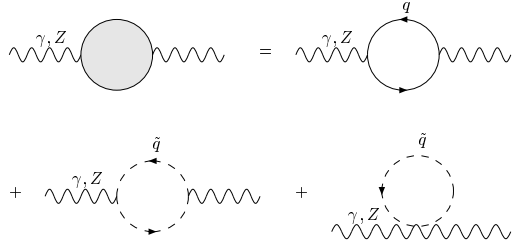,height=4cm,width=0.5\textwidth}}}
\caption{Top and bottom quark and squark contributions to the
unrenormalized $Z$ and $\gamma$ self-energies and to the $Z-\gamma$ mixing.} 
\label{ZGse} 
\end{figure} 
\begin{figure}
\centerline{\protect\hbox{\psfig{file=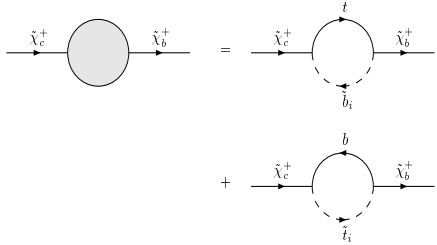,height=4cm,width=0.4\textwidth}}}
\caption{Feynman diagrams contributing to the unrenormalized
chargino two--point functions (self--energies and mixing).} 
\label{fig:cha2point} 
\end{figure}
\begin{figure}
\centerline{\protect\hbox{\psfig{file=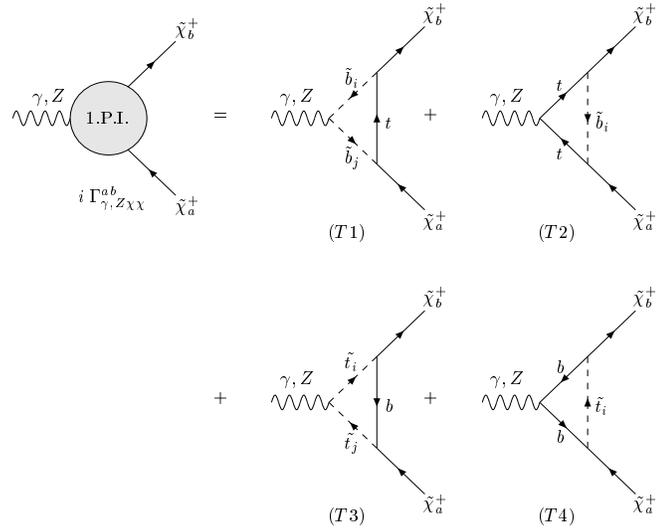,height=7cm,width=0.55\textwidth}}}
\caption{Unrenormalized one--particle irreducible triangular diagrams
contributing to renormalization of the $Z\tilde\chi^+\tilde\chi^-$ and 
$\gamma\tilde\chi^+\tilde\chi^-$ vertices.} 
\label{Tri1PI} 
\end{figure}

 
One-loop diagrams contributing to the gauge boson self energies and 
$Z-\gamma$ mixing can be seen in Fig.~\ref{ZGse}. The first diagram 
corresponds to quarks and the second two diagrams correspond to squarks.
Similarly, in Fig.~\ref{fig:cha2point} we display the diagrams contributing
to chargino self energies and mixing. They involve top--sbottom and
bottom--stop loops. Finally, 1PI triangular diagrams are shown in 
Fig.~\ref{Tri1PI}. All diagrams in Figs.~\ref{ZGse} to \ref{Tri1PI} have 
been calculated \cite{chargRC} in terms of Passarino--Veltman's 
functions \cite{VP}. We have used the $\overline{DR}$ scheme, which in our
approximation is completely equivalent to the $\overline{MS}$, and taken 
$Q=m_Z$ as the subtraction point.

\begin{figure}
\centerline{\protect\hbox{\psfig{file=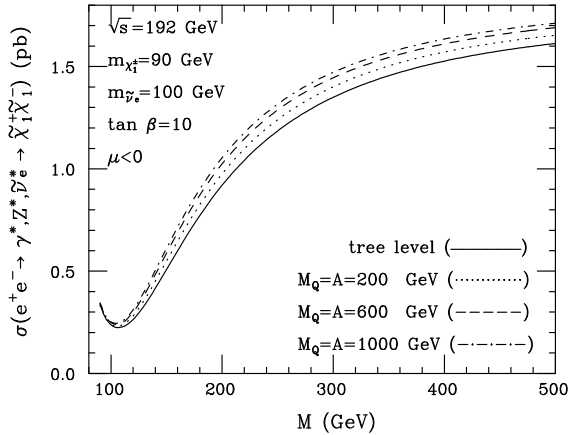,height=7cm,width=0.6\textwidth,angle=90}}}
\caption{One--loop and tree level chargino production cross section as a
function of the $SU(2)$ gaugino mass $M$, for 192 GeV center of mass energy.} 
\label{fig:c192m}
\end{figure}

In Fig.~\ref{fig:c192m} we have plot the tree--level and renormalized 
one--loop total production cross section of a pair of light charginos
as a function of the gaugino mass $M$, while keeping constant the
chargino mass $m_{\chi^{\pm}_1}=90$ GeV, the sneutrino mass 
$m_{\tilde\nu_e}=100$ GeV, and $\tan\beta=10$. 
We consider the case $\mu<0$ and a center of mass energy $\sqrt{s}=192$ 
GeV, relevant for LEP2. The tree level cross section decreases from 
1.6 pb. when $M=500$ GeV to a minimum of 0.22 pb. at around $M=105$ GeV, 
and grows again up to 0.34 pb. at $M=90$ GeV. Below this value of the 
gaugino mass $M$ there is no solution for $\mu<0$ which gives 
$m_{\chi^{\pm}_1}=90$ GeV. Radiative corrections to this cross section 
are parametrized by the squark soft masses which we take degenerate 
$M_Q=M_U=M_D$, and by the trilinear soft mass parameters $A\equiv A_U=A_D$, 
also taken degenerate. This choice is done at the weak scale and it is 
made for simplicity. Three radiatively corrected curves are presented
given by $M_Q=A=200$ GeV (dots), $M_Q=A=600$ GeV (dashes), and $M_Q=A=1$ 
TeV (dotdashes). 

We observe from Fig.~\ref{fig:c192m} that radiative corrections are 
positive and grow with the squark mass parameters. For $M$ close to
90 GeV the corrections are only of a few percent, but they grow fast
until a maximum of $21\%$ at $M=140$ GeV. For larger values of the gaugino 
mass, the corrections slowly decrease until they reach the value $6\%$
at $M=500$ GeV. A logarithmic growth of quantum corrections with the 
squark mass parameters is observed, as it should be. It is worth pointing 
out that the value of $\mu$ is not constant along the curves because it 
is fixed by the constant value of the chargino mass $m_{\tilde\chi^{\pm}_1}$.

In summary, if charginos are discovered much information can be learned
from the measurements of the total production cross section and masses.
Nevertheless it is essential to have a precise theoretical calculation
of these observables. In this sense, one--loop radiative corrections
must be included. We have found that for LEP2 energies they are typically 
$10\%$ to $15\%$ and can reach up to $30\%$ if the squark masses are 
equal to 1 TeV. 

\section*{Acknowledgments}
The authors are grateful to PPARC for partial support under contract 
no. GR/K55738. M.A.D. was also supported by a postdoctoral grant from 
Ministerio de Educaci\'on y Ciencias, by DGICYT grant PB95-1077 and by 
the EEC under the TMR contract ERBFMRX-CT96-0090.

\end{document}